\documentclass[12pt,preprint]{aastex}

\shorttitle{ROSAT HRI IXO Catalog}
\shortauthors{Colbert et al.}

\begin{document}

\title{A Catalog of Candidate Intermediate-luminosity X-ray Objects}

\author{E. J. M. Colbert and A. F. Ptak}
\affil{
Department of Physics and Astronomy, 
Johns Hopkins University, 
Baltimore, MD~~21218
}

\begin{abstract}

\noindent
{\bf See {\it http://www.xassist.org/ixocat\_hri} for figures and future updates
of this catalog. NOTE -- some IXO numbers have changed since the original
astro-ph submission.  This paper is now accepted into the ApJ Supplements,
and Tables 1 and 2 comprise version 2.0 of our ROSAT HRI IXO catalog.}
\\
\\
ROSAT, and now {\it Chandra}, X-ray images allow studies of
{\it extranuclear} X-ray point
sources in galaxies other than our own.  X-ray observations of normal
galaxies with ROSAT and Chandra have revealed that off-nuclear, compact,
Intermediate-luminosity (L$_X[2$-$10~{\rm keV}] \ge$10$^{39.0}$ erg~s$^{-1}$)
X-ray Objects (IXOs, a.k.a. ULXs [Ultraluminous X-ray sources]) 
are quite common.  Here we present a 
catalog and finding charts for 87 IXOs in 54 galaxies, derived from all of 
the ROSAT HRI imaging data for galaxies with 
$cz \le$ 5000 km~s$^{-1}$ from the
Third Reference Catalog of Bright Galaxies (RC3).  We have defined the
cutoff L$_X$ for IXOs so that it is well above the Eddington luminosity
of a 1.4 M$_\odot$ black hole (10$^{38.3}$ erg~s$^{-1}$), so as not to
confuse IXOs with ``normal'' black hole X-ray binaries.  This catalog
is intended to provide a baseline for follow-up work with {\it Chandra}
and XMM,
and with space- and ground-based survey work at wavelengths other than 
X-ray.  We demonstrate that elliptical galaxies with IXOs have a larger 
number of IXOs per galaxy than non-elliptical galaxies with IXOs, and 
note that they are not likely to be merely high-mass X-ray binaries with
beamed X-ray emission, as may be the case for IXOs in starburst galaxies.
Approximately half of the IXOs with multiple observations show X-ray
variability,
and many (19) of the IXOs have faint optical counterparts in DSS optical
B-band images.  Follow-up observations of these objects should be
helpful in identifying their nature.
\end{abstract}

\keywords{
catalogs,
galaxies: general,
X-rays: binaries,
X-rays: galaxies
}

\catcode`\@=11
\def\gapprox{\mathrel{\mathpalette\@versim>}}
\def\lapprox{\mathrel{\mathpalette\@versim<}}
\def\@versim#1#2{\lower2.45pt\vbox{\baselineskip0pt\lineskip0.9pt
    \ialign{$\m@th#1\hfil##\hfil$\crcr#2\crcr\sim\crcr}}}
\catcode`\@=12

\section{Introduction}

In the early 1980s, surveys of {\it normal} galaxies with the Einstein
X-ray satellite revealed intermediate-luminosity (L$_X$ $\sim$
10$^{39}$$-$10$^{40}$ erg~s$^{-1}$) X-ray sources which were seemingly
located in the centers of spiral galaxies (cf. Fabbiano 1989).  
This was 
very interesting, since Seyfert nuclei (active galactic nuclei
[AGNs] in nearby spirals) are typically much more luminous (L$_X$
$\sim$ 10$^{42}$$-$10$^{44}$ erg~s$^{-1}$) and black hole X-ray
binaries (BH XRBs) are much less luminous (typically L$_X$ $\lapprox$
10$^{38}$ erg~s$^{-1}$).  It was not clear whether these intriguing
sources were underluminous accreting supermassive BHs, over-luminous
XRBs located near the galactic nucleus, or an entirely new type of
astrophysical object altogether (e.g., Colbert et al. 1995).  In the
1990s, ROSAT observations with the High Resolution Imager (HRI) showed
that these Intermediate-luminosity X-ray Objects 
(IXOs, a.k.a.  ``Ultraluminous X-ray sources,'' or ULXs) 
are compact
X-ray sources, and are quite common in the local Universe -- present in
the nuclear regions of $\sim$50\% of nearby normal galaxies observed
with the ROSAT HRI (Colbert \& Mushotzky 1999; Roberts \& Warwick
2000, hereafter RW2000).  Many IXOs were found to be offset from the optical nucleus (e.g.,
Colbert \& Mushotzky 1999).  

As an example, 
the ``near-nuclear'' IXO in NGC~1313
is actually displaced $\sim$1$'$ ($\sim$ 1.5 kpc) from the optical nucleus.
Such a displacement is not expected if the object is a supermassive BH
(M $\gapprox$ 10$^6$ M$\odot$), since dynamical friction would likely
cause the BH to sink to the center of the galaxy in a Hubble time (cf.
Tremaine, Ostriker \& Spitzer 1975).  For this reason, IXO BH masses
are estimated to be $\lapprox$10$^5$ M$\odot$.  
If the X-rays are emitted isotropically and the luminosities are sub-Eddington,
the mass of the
central object 
is required to be
$\gapprox$ 10 M$_\odot$.  
Therefore IXOs, if powered by accretion
and not beamed (cf. King et al. 2001 and Koerding, Falcke \& Markoff 2002
for discussions on
beaming), have {\it intermediate-mass black holes} (IMBHs).  If this is
true, then IXOs are extremely interesting objects since they are the only
tracers of this peculiar class of black holes.  Formation of IMBHs is 
not well understood -- for example, supernova explosions produce BHs
$\lapprox$ 20 M$\odot$.  The merging of smaller mass black holes has
been proposed as a formation mechanism for IMBHs 
(e.g. Taniguchi et al. 2000, Miller \& Hamilton 2002).
Continued observational and theoretical work on IXOs will hopefully help 
unravel their mysterious nature.

There are currently many important, unanswered questions about IXOs.  For
example:  Do they really contain IMBHs?  How did IXOs form?  
Are they binary systems?  Are IXOs merely ``normal'' stellar-mass BH XRBs
with beamed X-ray emission?  
How are the BHs
fueled?  Are IXOs more numerous in galaxies with high star formation
rates?  If so, is that because there are more seed black holes to merge into
IMBHs, more fuel, or
for some other reason?  
Are IXOs in elliptical galaxies physically different types of objects than
IXOs in spiral galaxies?

Since IXOs are usually found displaced from the center of the galaxy, much of
the current work on IXOs stems from comparisons of their properties with 
those of ``normal'' BH XRBs
and ``micro-quasars'' (e.g., Colbert et al. 1999, Makishima et al. 2000, 
Roberts et al. 2002, and Koerding, Falcke \& Markoff 2002).  Reviews of
the properties of BH XRBs and micro-quasars can be found in Tanaka et al. 
(1995) and Mirabel (1998), respectively.

Thus far, most of the effort to understand IXOs has been focussed on a few
well-known objects (e.g., the M82 IXO, N5204 X-1, NGC 3628 X-1).
Comparatively fewer studies have been done at wavelengths other 
than X-ray, and this
is partly due to a lack of a complete catalog of IXOs.
Optical surveys will allow studies of the properties of
the accreting gas (star or dense gas
cloud) immediately around the BH, and of the local galactic environment
(e.g. globular clusters, gaseous nebulae, young star clusters).
This
could then provide clues to how IXOs might have formed, or at least how they
might have become ``active'' X-ray emitters.
Here we present a catalog of 87 IXO candidates that were carefully selected from
all of the public ROSAT HRI images of galaxies from the
Third Reference Catalog of Bright Galaxies (hereafter RC3,
de Vaucouleurs et al. 1991).  

\section{Data Reduction}

Our IXO catalog is based on a cross-correlation between all of the
X-ray point sources in ROSAT HRI images and galaxies listed in v3.9b of RC3.

Distances to RC3 galaxies were
extracted from galaxy distance catalogs (see references listed in Table~1), or 
were computed
from the recessional velocity $cz$ using H$_0 =$ 75 km~s$^{-1}$~Mpc$^{-1}$,
if $cz >$ 1000 km~s$^{-1}$.  Galaxies with unlisted values of $cz$
and no known distances were rejected.  We also rejected very distant galaxies,
by rejecting galaxies with $cz >$ 5000  km~s$^{-1}$ (66.7 Mpc for  
H$_0 =$ 75 km~s$^{-1}$~Mpc$^{-1}$).
After selection, the number of RC3 galaxies in our sample was 9999.
We have revised the distance to NGC~3031 (M81) from 1.4 Mpc (Tully 1998) to
the more accurate and much larger Cepheid distance of 3.55 Mpc (Paturel
et al. 2002).  We did not find any other of our galaxy distances from Tully 
(1998) that were highly discrepent with recent Cepheid distances.

A list of 5465
ROSAT HRI datasets was compiled, based on all of the public HRI data available
on the FTP archive at NASA/GSFC on February 18, 2001.  
The positions of the pointing centers for each HRI image were 
cross-correlated with the
positions of each of the 9999
galaxies listed in our RC3 galaxy sample.
Since the HRI FOV is
36$'$ in diameter,
we selected HRI images for reduction if they had pointing centers 
offset $\le$ 40$'$ from any of the 9999 accepted RC3 galaxies.  After selection,
our data sample consisted of 1883
unique ROSAT HRI datasets.

Reduction of the ROSAT HRI data was done using the semi-automated X-ray data
reduction package XASSIST (http://www.xassist.org).  The 
first step of the reduction
process consists of finding point sources in the images.  
Each of the raw 1883 HRI 
images was analyzed using a wavelet point source detection algorithm 
(the ``detect'' feature in XASSIST) to construct a list of potential point 
sources.  After the detection algorithm, a more rigorous spatial fitting 
test was performed for each point
source, which consisted of fitting a 2D Gaussian model and a 
sloping background model
to the image at each of the source positions.  
The resulting sources were then flagged
by XASSIST as being point-like, extended, or asymmetric.  Only point-like 
sources with ``significance'' (S/N) greater than 3.0 were kept.  The final
list consisted of 4281 point sources.  These sources were then cross-correlated
with RC3, and sources within 10$"$ of each other were identified as the same
source.  
Count rates were then converted to 2$-$10 keV fluxes using PIMMS and an
assumed power-law spectrum with $\Gamma=$1.7 (see also Table 1 notes).  As
noted by Strickland et al. (2001) and Roberts et al. (2002), many Chandra
spectra of IXOs are consistent with this assumed spectral model.  Isotropic
luminosities were then calculated from the fluxes, assuming that the IXOs
were located at the distance of the corresponding galaxy.
A filtered list was then made of all sources with 
L$_X(2-10~{\rm keV}) \ge 10^{39}$ erg~s$^{-1}$ and positions 
within two R$_{25}$ 
radii of an RC3 galaxy.
The optical fields around the potential IXOs were then individually inspected
and sources that were obvious mis-associations (e.g., nuclei of other nearby
galaxies, and sources located far from edge-on disk galaxies, perpendicular to 
the disk) were rejected.
This list was then cross-correlated with the NED and Simbad databases to 
further filter out QSOs and X-ray supernova.  
Our final catalog of 87
IXOs is listed in Table 1.

\section{Finding Charts}

In Figures 1$-$15, we show contour plots 
of the ROSAT HRI X-ray emission from the 87 IXOs,
overlayed on B-band Digital Sky Survey (DSS) images from 
STScI.  The datasets used for the contour plot are listed in Table 2.

\section{Comments on Individual Objects}

\subsection{IXO 1 and IXO 2}

Both of these objects are located near the elliptical galaxy NGC~720
(Figure 1).  They have faint optical counterparts in the DSS images, but
the counterparts are not listed in either the NED or the Simbad databases.
See also Fabbiano, Trinchieri \& Kim (1992, hereafter FKT92).

\subsection{IXO 3}

This object is located at the outer edge of the disk in the edge-on spiral
galaxy NGC~891 (Figure 1).  It is also known as NGC~891 X-5 
(RW2000).
Count rates from two separate HRI observations of IXO~3 show $\sim$50\%
variability (Table 2).
IXO~3 was detected with ASCA (cf. Ueda et al. 2001), but was not detected
with the ROSAT PSPC (e.g., Read, Ponman \& Strickland 1997, hereafter RPS97).

\subsection{IXO 4}

IXO 4 is located in the disk of the face-on spiral galaxy NGC~1042 (Figure 1).
It is quite luminous, with an estimated 2$-$10 keV luminosity of 
1 $\times$ 10$^{40}$ erg~s$^{-1}$.
See also FKT92 and Ueda et al. (2001).

\subsection{IXO 5}

IXO 5 is also located in the disk of a face-on spiral galaxy (NGC~1073, see
Figure 1).
There are two other bright X-ray point sources positioned within this galaxy.
These other two sources are coincident with known QSOs (see Figure 2).

\subsection{IXO 6}

This object is positioned in the outer disk (or halo) of the early-type
galaxy NGC~1291 (Figure 2).  See also RPS97.

\subsection{IXOs 7 and 8}

IXO 7 is a well-studied X-ray source (Fabbiano \& Trinchieri 1987, Petre
et al. 1994, Colbert
et al. 1995, Miller et al. 1998, Schlegel et al. 2000b).  Its ASCA spectra
show spectral variability (e.g., Colbert \& Mushotzky 1999), similar to
that of a black-hole XRB.  It is located at the northern end of the bar, 
$\sim$1 kpc from the center of the face-on spiral galaxy NGC~1313 (Figure 2).
IXO~7 is commonly referred to as NGC~1313~X-1.

IXO 8 (NGC~1313 X-2) is positioned in the outer disk of NGC~1313, $\sim$7'
south from X-1 (Figure 2).  See Makishima et al. (2000) and Mizuno et al. 
(2001) for ASCA spectral analyses of this source.

The count rates from numerous HRI observations of these two IXOs show that they
are variable by a factor of $\sim$2 (Table 2).

\subsection{IXOs 9 $-$ 14}

IXOs 9$-$14 are positioned in the outskirts of the elliptical galaxy NGC~1316
(Fornax~A; see Figure 2).  IXOs 9, 10 and 14 have faint optical
counterparts in the DSS image, but there are no
catalogued sources in NED or Simbad at those positions.  
Kim, Fabbiano, \& Mackie (1998) have published an extensive study of the ROSAT 
X-ray sources in Fornax~A.
See also FKT92.

\subsection{IXOs 15 and 16}

These two objects are positioned very near each other, in the outer disk
of the spiral galaxy NGC~1365; see Figure 3).  Possible optical counterparts
are visible in the DSS images.  See Turner, Urry \& Mushotzky (1993) and 
Singh (1999) for more detailed discussions of these X-ray objects.
See also FKT92.

\subsection{IXOs 17, 18, 19 and 20}

These four objects are positioned near the elliptical galaxy NGC~1399
(see Figure 3).  IXOs 17 and 18 are near the (extended) nuclear X-ray 
source and are not well resolved by the HRI.
IXO 19 is located much further out in the halo, and is coincident with a 
faint optical counterpart seen in the DSS image.  However, there is no
catalogued entry for the counterpart in NED or Simbad.  IXO 20 is also
coincident with an optical DSS counterpart.
IXO~19 shows variability by a factor of $\sim$3, while IXO~20 shows $\sim$50\%
variability between two HRI observations (Table 2).
See also Jones et al. (1997).

\subsection{IXO 21}

IXO 21 is positioned in the outer regions of the irregular galaxy NGC~1427A
(Figure 3).  See also Jones et al. (1997).

\subsection{IXO 22}

IXO 22 is positioned in the spiral arms of the nearby, face-on spiral galaxy
IC~342 (Figure 3).  It is commonly known as IC~342 X-1.  It is coincident with
an optical counterpart seen in the DSS image, but NED and Simbad have no
catalog entry for the counterpart. IC~342 has a low Galactic latitude 
($b \approx$10$^\circ$), so the counterpart could be a Galactic star.
ASCA studies of IXO~22 (e.g., Makishima et al. 2000, Kubota et al. 2001, 
Mizuno et al. 2001) show that this source is quite variable.  See also 
Fabbiano \& Trinchieri (1987), Bregman, Cox \& Tomisaka (1993), 
Colbert \& Muzhotzky (1999), RW2000, and Lira et al. (2000).

\subsection{IXO 23}

This object is positioned in the outskirts of the elliptical galaxy NGC~1553 
(Figure 4).
See Blanton, Sarazin \& Irwin (2001) for a discussion of the Chandra X-ray
point sources in NGC~1553.

\subsection{IXO 24}

IXO 24 is located near the center of
the face-on spiral galaxy
NGC~1566 (Figure 4).
This source is surrounded by strong extended (diffuse?) X-ray emission from
the disk of NGC~1566.

\subsection{IXOs 25, 26 and 27}

These three objects are located in the spiral arms of the galaxy NGC~1672 (see 
Figure 4).  See Brandt, Halpern \& Iwasawa (1996) and De~Naray et al. (2000)
for more detailed
discussions of these X-ray sources.  Count rates from two HRI observations of
IXO~26 show variation by a factor $\sim$2.5.

\subsection{IXO 28}

This IXO is located in the disk of the spiral galaxy NGC~1792 (Figure 4).
See Dahlem et al. (1994) for a ROSAT analysis of the point sources in NGC~1792.

\subsection{IXO 29}

IXO 29 is located in the outskirts of the spiral galaxy NGC~1961 (Figure 5).
See also FKT92, Pence \& Rots (1997), and RW2000.

\subsection{IXO 30}

IXO 30, located $\sim$2$'$ west of the center of the Seyfert galaxy Mrk~3
(Figure 5),
is discussed by Turner, Urry \& Mushotzky (1993) and Morse et al. 
(1995).

\subsection{IXO 31}

This IXO in the dwarf galaxy Holmberg~II (Figure 5) is well studied in X-rays
(e.g., Colbert \& Mushotzky 1999, Zezas, Georgantopoulos, \& Ward 1999, and
Miyaji et al. 2001).  
It is located in the middle of an H~II region complex, and is also coincident
with a luminous radio source (cf. Miyaji et al. 2001).

\subsection{IXOs 32 and 33}

These two objects are positioned in the outskirts of the early-type spiral
galaxy NGC~2775 (Figure 5).  They are both coincident with faint DSS sources,
although the optical counterparts are not listed in NED or Simbad.
See also RW2000.

\subsection{IXO 34}

IXO 34 is a well-known X-ray source (M81 X-9) near the dwarf galaxy
Holmberg~IX, a companion to NGC~3031 (M81; see Figure 6).  It has been 
hypothesized to be a superbubble (Miller 1995), a background quasar 
(Ishisaki et al. 1996; Ezoe et al. 2002), and an accreting intermediate-mass
black hole (Wang 2002).  See La~Perola et al. (2001) for a discussion of
this X-ray source.  IXO~34 is highly variable.  As can be seen from the HRI
count rates listed in Table 2, it shows variation in count rate by a factor
of $\sim$3.
See also Fabbiano (1988), FKT92, Makishima et al. (2000), RW2000, 
Lira et al. (2000), Immler \& Wang (2001), and Mizuno et al. (2001).

\subsection{IXO 35}

IXO 35 is located $\sim$2$'$ north of the elliptical galaxy NGC~3226 (see 
Figure 6).  There is a possible optical counterpart to IXO 35 in the DSS
image, but it is not cataloged in NED or Simbad.
See also RW2000.

\subsection{IXO 36}

This object is located $\sim$2.5$'$ to the west of the center of the galaxy
merger system NGC~3256 (Figure 6).

\subsection{IXO 37}

This IXO is located in the outskirts of the elliptical galaxy IC~2597, which 
is in the Hydra cluster (see Figure 6).  There is a faint optical sources
near the IXO in the DSS image.  The IXO is also positioned $\sim$10$"$ from
a radio source (Andernach et al. 1988).

\subsection{IXO 38}

IXO 38 is located $\sim$1.5$'$ to the northwest of the center of the spiral
galaxy NGC~3310 (Figure 7).  See Zezas, Georgantopoulos, \& Ward (1998) for
more details of this X-ray source.
See also RW2000.

\subsection{IXO 39}

IXO 39 is located near the nucleus of the edge-on starburst galaxy NGC~3628 
(Figure 7).  It is also known as NGC~3628 X-1.  See 
Dahlem, Heckman \& Fabbiano (1995) and
Strickland et al. (2001) for discussions of this well-known IXO.  See also
Bregman \& Glassgold (1982), FKT92, Yaqoob et al. (1995), RPS97, 
Dahlem, Weaver \& Heckman (1998), and RW2000.

\subsection{IXO 40}

This X-ray object is positioned $\sim$5$'$ to the north of the elliptical
galaxy NGC~3923 (Figure 7).  It is coincident with an optical counterpart
in the DSS image, but the counterpart is not listed in NED or Simbad.

\subsection{IXO 41}

IXO 41 is positioned $\sim$0.5$'$ to the west from the Irregular galaxy
UGC~7009 (Figure 7).

\subsection{IXO 42}

IXO 42 is located in the disk of the edge-on spiral galaxy NGC~4088 (Figure 8).
It is also known as NGC~4088 X-1 (RW2000).

\subsection{IXOs 43 and 44}

These two IXOs are positioned $\sim$4$'$ and $\sim$2$'$ west, respectively,
of the Seyfert nucleus in NGC~4151 (see Figure 8).  See 
Warwick, Done, \& Smith (1995) for a more detailed discussions of these two
objects.
See also RW2000.

\subsection{IXO 45}

IXO 45 is positioned $\sim$3$'$ to the southeast of the elliptical galaxy
NGC~4203 (Figure 8).
It is also positioned near the QSO Ton~1480.
See also RW2000.

\subsection{IXO 46}

This X-ray object is located in the disk of the spiral galaxy NGC~4254 
(Figure 8).  There is a bright optical counterpart visible in the DSS image, 
but it is not catalogued in NED or Simbad.

\subsection{IXOs 47, 48 and 49}

These three IXOs are located in the disk of the nearby spiral galaxy 
NGC~4321 (M100; see Figure 9).  IXOs 47 and 48 are sources 
H13 and H14 (respectively) in the ROSAT HRI study by 
Immler, Pietsch, \& Aschenbach (1998).
See also RW2000.

\subsection{IXOs 50 and 51}

IXO 50 is a slightly resolved X-ray source located just east of the nucleus
of the elliptical galaxy NGC~4374 (see Figure 9).  IXO~51 is positioned
$\sim$3.5$'$ to the northeast of NGC~4374.
There is also an X-ray bright QSO near NGC~4374. 
See also FKT92 and Finoguenov \& Jones (2001).

\subsection{IXO 52}

IXO 52 is positioned $\sim$2$'$ north of the elliptical 
galaxy NGC~4393 (Figure 9).  There is a faint optical counterpart visible
in the DSS image.

\subsection{IXO 53}

This IXO is located east of the dwarf
galaxy NGC~4395 (see Figure 10).  It is also known as NGC~4395 X-2 
(Lira, Lawrence, \& Johnson 2000).
See also Iwasawa et al. (2000) and RW2000.

\subsection{IXO 54}

IXO 54 is positioned $\sim$3$'$ northeast of the edge-on spiral galaxy 
NGC~4438 (Figure 10).
See also RW2000.

\subsection{IXOs 55 $-$ 61}

These seven IXOs are positioned in the halo of the elliptical galaxy 
NGC~4472 (Figures 10 and 11).
See also Irwin \& Sarazin (1996).

\subsection{IXO 62}

IXO 62 is located in the galaxy NGC 4485, which is a small companion to the
spiral galaxy NGC~4490 (see Figure 11).  
Several HRI observations of IXO~64 show that it is variable by a factor of
$\sim$2 (Table 2).  
There are several extended X-ray sources in the disk of NGC~4490 that have been
resolved into individual point sources with Chandra (Roberts et al. 2002).
See also Colbert \& Mushotzky (1999) and RW2000.

\subsection{IXOs 63 and 64}

IXOs 63 and 64 are positioned in the halo of the elliptical galaxy NGC~4552
(see Figure 11).

\subsection{IXO 65 and 66}

These two IXOs are positioned in the spiral galaxy NGC~4559 (see Figure 11).
IXO 65 is located $\sim$3$^\prime$ southwest of the galaxy center, while IXO~66
is located near the galaxy center.  IXO~65 shows X-ray variability (Table 2)
and has an optical counterpart in the DSS image.  See also Vogler, Pietsch \&
Bertoldi (1997) and RW2000.

\subsection{IXO 67}

IXO 67 is positioned in the halo of the edge-on spiral galaxy NGC~4565, 
$\sim$1$'$ from the nucleus (see Figure 12).  
See Vogler, Pietsch, \& Kahabka (1996) and Wu et al. (2002) 
for more discussions of IXO 66.
See also Mizuno et al. (1999) and Makishima et al. (2000).

\subsection{IXO 68}

IXO 68 is positioned in the disk of the edge-on
spiral galaxy NGC~4631 (see Figure 12).  See Vogler \& Pietsch (1996) for
a more detailed discussion of the X-ray properties of these IXOs.
Count rate measurements from several HRI observations of IXO~68 show that
it is variable by a factor of $\sim$2.
See also RPS97, Dahlem, Weaver \& Heckman (1998), and RW2000.

\subsection{IXOs 69 $-$ 71}

These three IXOs are positioned in the halo of the elliptical galaxy NGC~4649
(M60; see Figure 12).
Count rate measurements from two HRI observations of IXO~73 show variability
of $\sim$50\% (Table 2). 
See also FKT92.

\subsection{IXOs 72 and 73}

IXOs 72 and 73 are located in the irregular galaxy NGC~4861 (see Figure 12).
IXO~73 varied by a factor of $\sim$5 between two separate HRI observations
(Table 2).
See also Stevens \& Strickland (1998).

\subsection{IXO 74}

IXO 74 is positioned in the outer disk of the spiral galaxy NGC~5055 
(Figure 13).  It is also known as NGC~5055 X-2 (RW2000).
This X-ray source is quite variable.  It shows a factor of $\sim$2 variability
between two HRI observations (Table 2).
See also RPS97.

\subsection{IXOs 75 and 76}

These two IXOs are positioned in the halo/bulge of NGC~5128 (Cen~A; see
Figure 13).  IXO~75 is coincident with a bright optical counterpart in the 
DSS image, but the counterpart is not catalogued in NED or Simbad.  See
Turner et al. (1997) for a more detailed discussion of these two X-ray 
sources.
Count rates from multiple HRI observations of these two sources
show variability by a factor of $\sim$2 (Table 2).  See Steinle, Dennerl \&
Englhauser (200) for a detailed ROSAT analysis of these point sources.

\subsection{IXO 77}

IXO 77 is also known as NGC~5204 X-1 (e.g., Lira et al. 2000, 
Roberts et al. 2001).  It is positioned near an H~II region complex in the
irregular galaxy NGC~5204 (Figure 13).
See also Colbert \& Mushotzky (1999) and RW2000.

\subsection{IXOs 78 $-$ 81}

These four IXOs are located in the spiral arms of NGC~5194 
(M51; see Figure 13).
See Ehle, Pietsch, \& Beck (1995) and RPS97 for
more detailed discussions of these IXOs.
From the measured count rates from multiple HRI observations of these IXOs
(Table 2), we find that IXO~79 varies by a factor of $\sim$3, while IXO~81 
shows $\sim$50\% variability.
See also Palumbo et al. (1985), Fabbiano \& Trinchieri (1987), FKT92, 
Marston et al. (1995), and RW2000.

\subsection{IXO 82}

This IXO is located in the outer disk of NGC~5236 (M83; see Figure 14).  See
Ehle et al. (1998) and Immler et al. (1999)
for detailed X-ray studies of M83.
The measured count rates for two HRI observations of this sources differ
by a factor of $\sim$2 (Table 2).
See also Lira et al. (2000).

\subsection{IXO 83}

IXO 83 is located in the outer disk of NGC~5457 (M101; see Figure 14).  See
RPS97 and Wang, Immler \& Pietsch (1999) for ROSAT
analyses of M101.
IXO~83 shows $\sim$50\% variability between four difference HRI observations
(Table 2).
See also RW2000 and Lira et al. (2000).

\subsection{IXO 84}

This IXO is positioned near NGC~5774 and NGC~5775 in the NGC~5775
group (see Figure 14).
See also RW2000.

\subsection{IXO 85}

IXO 85 is located near the center of the face-on spiral galaxy NGC~6946
(see Figure 14).
There are many very luminous X-ray SNRs in NGC~6946, although IXO~85 is not
classified as such.  
See Schlegel, Blair \& Fesen (2000a) for details.
It shows variability by a factor of $\sim$2.5 between
two different HRI observations (Table 2).
See also Colbert \& Mushotkzy (1999), RW2000, and Lira et al. (2000).

\subsection{IXO 86}

IXO 86 is located in the outer disk of the spiral galaxy NGC~7314 (see 
Figure 15).  See also Radecke (1997).

\subsection{IXO 87}

This X-ray object is located in the disk of the Seyfert galaxy NGC~7590
(see Figure 15).  See also Ueda et al. (2001).

\section{Discussion and Summary}

We have found 87 IXOs in 54 different galaxies.  We find the following 
statistical results from perusal of the optical images and from the
information listed in Table 1.

IXOs are found both in
spiral and elliptical galaxies, although elliptical galaxies have a higher
number of IXOs per galaxy (mean of 2.3 per galaxy for T-type $<$ 0, {\it vs.}
1.6 for T-type $\ge$ 0).  For example, the elliptical galaxy
NGC~1316 (Fornax~A) has an astounding 7 IXOs in its halo.

Nineteen of the 87 IXOs are positioned near faint optical sources in the DSS
images, and many more are located in the disks of spiral galaxies, or are known
from more recent optical observations to have counterparts (e.g., NGC 5204 X-1,
Roberts et al. 2001; see also Pakull et al. 2002).
Follow-up studies of these potential optical counterparts using
{\it Chandra} X-ray positions would be very useful for understanding the
nature of IXOs.  IXOs are, in general, quite variable:  51\% (20 of 39) of the
objects with two or more HRI observations show $\gapprox$50\% variability 
between the observations.

We note here that the IXOs in this catalog are selected strictly by their
X-ray luminosity and proximity to nearby RC3 galaxies.
It is possible that some of the objects listed in the catalog are background
quasars, as can be deduced from the number of QSOs seen in the finding
charts (Figures 1 $-$ 15).  
Thus, this catalog should be viewed as a list of IXO {\it candidates}, 
especially for objects positioned at large projected radii from the galaxy
nucleus.  On the other hand, many of the X-ray sources, such as those in 
NGC~1313 and NGC~5194, for example, are very likely associated with their
host galaxies.

Although an overabundance of IXOs are often found in starburst galaxies (e.g.
Roberts et al. 2002), we have shown here that there is a significant population
(35 of the 87) of IXOs in the halos of elliptical galaxies, where there is 
no recent star formation.  This suggests that IXOs in elliptical galaxies may
not be merely young, high-mass XRBs with mildly beamed X-ray emission 
(e.g., King et al. 2001).  A more extensive analysis of the 
properties 
of these ROSAT IXOs, together with a study of their host galaxies, will
be presented in a later work.

The current version and future versions of this catalog are available on
the World Wide Web (http://www.xassist.org/ixocat\_hri).

\acknowledgments

The Digitized Sky Survey (DSS) was produced at the Space Telescope Science
Institute under U.S. Government grant NAG W-2166. This research has
made use of the NASA/IPAC Extragalactic Database (NED) which is
operated by the Jet Propulsion Laboratory, California Institute of
Technology, under contract with the National Aeronautics and Space
Administration. This research has also made use of the SIMBAD database,
operated at CDS, Strasbourg, France.  EJMC acknowledges support from NASA
LTSA grant NAG 59878.  We thank the anonymous referee and Manfred Pakull
for useful comments.

\clearpage

%\begin{figure}
%\epsscale{1.0}
%\plotone{f01.eps}
%\caption{}
%\end{figure}
%
%\clearpage
%
%\begin{figure}
%\epsscale{1.0}
%\plotone{f02.eps}
%\caption{}
%\end{figure}
%
%\clearpage
%
%\begin{figure}
%\epsscale{1.0}
%\plotone{f03.eps}
%\caption{}
%\end{figure}
%
%\clearpage
%
%\begin{figure}
%\epsscale{1.0}
%\plotone{f04.eps}
%\caption{}
%\end{figure}
%
%\clearpage
%
%\begin{figure}
%\epsscale{1.0}
%\plotone{f05.eps}
%\caption{}
%\end{figure}
%
%\clearpage
%
%\begin{figure}
%\epsscale{1.0}
%\plotone{f06.eps}
%\caption{}
%\end{figure}
%
%\clearpage
%
%\begin{figure}
%\epsscale{1.0}
%\plotone{f07.eps}
%\caption{}
%\end{figure}
%
%\clearpage
%
%\begin{figure}
%\epsscale{1.0}
%\plotone{f08.eps}
%\caption{}
%\end{figure}
%
%\clearpage
%
%\begin{figure}
%\epsscale{1.0}
%\plotone{f09.eps}
%\caption{}
%\end{figure}
%
%\clearpage
%
%\begin{figure}
%\epsscale{1.0}
%\plotone{f10.eps}
%\caption{}
%\end{figure}
%
%\clearpage
%
%\begin{figure}
%\epsscale{1.0}
%\plotone{f11.eps}
%\caption{}
%\end{figure}
%
%\clearpage
%
%\begin{figure}
%\epsscale{1.0}
%\plotone{f12.eps}
%\caption{}
%\end{figure}
%
%\clearpage
%
%\begin{figure}
%\epsscale{1.0}
%\plotone{f13.eps}
%\caption{}
%\end{figure}
%
%\clearpage
%
%\begin{figure}
%\epsscale{1.0}
%\plotone{f14.eps}
%\caption{}
%\end{figure}
%
%\clearpage
%
%\begin{figure}
%\epsscale{1.0}
%\plotone{f15.eps}
%\caption{}
%\end{figure}
%
%\clearpage
%
%
%
%
%
%%

\clearpage

\begin{deluxetable}{rrrclcccrlrr}
\tabletypesize{\scriptsize}
\tablecaption{ROSAT HRI IXO catalog\label{tab1}}
\tablewidth{0pt}
\tablehead{
\colhead{IXO} & \colhead{R.A.}   & \colhead{Dec.}  & 
\colhead{L$_X$} &
\multicolumn{6}{c}{ 
--------------------------- 
Host Galaxy Properties 
--------------------------- 
} &
\colhead{R} & \colhead{R$_{max}$} \\
\colhead{} & \multicolumn{2}{c}{(J2000)} &
\colhead{} &
\colhead{Galaxy Name} & \colhead{T-type} & \colhead{AGN} & \colhead{log(FIR/B)} & \multicolumn{1}{c}{D} &
\colhead{Ref.} &
\colhead{} & \colhead{} \\
\colhead{} & \colhead{} & \colhead{} & \colhead{} &
\colhead{} & \colhead{} & \colhead{Type} & \colhead{} & \colhead{(Mpc)} &
\colhead{} &
\colhead{(kpc)} & \colhead{(kpc)} \\
\colhead{(1)} & \colhead{(2)} & \colhead{(3)} & \colhead{(4)} &
\colhead{(5)} & \colhead{(6)} & \colhead{(7)} & \colhead{(8)} & \colhead{(9)} &
\colhead{(10)} & \colhead{(11)} & \colhead{(12)} \\ 
}
\startdata
 1 & 01 52 49.7 & $-$13 42 11 & 39.1 & NGC   720      &  $-$5.0 &      &        & 22.88 & 1 &  22.5 & 31.1 \\
 2 & 01 53 02.9 & $-$13 46 53 & 39.7 & NGC   720      &  $-$5.0 &      &        & 22.88 & 1 & 17.3 & 31.1 \\
 3 & 02 22 45.8 & $+$42 25 55 & 39.2 & NGC   891      &  $+$3.0 &      & $+$1.0 &  9.60 & 2 & 15.7 & 37.7 \\
 4 & 02 40 25.8 & $-$08 24 30 & 40.1 & NGC  1042      &  $+$6.0 &      & $+$0.0 & 18.31 & 1 &  8.5 & 24.9 \\
 5 & 02 43 38.3 & $+$01 24 13 & 39.1 & NGC  1073      &  $+$5.0 &      & $-$0.1 & 16.15 & 1 &  8.0 & 23.0 \\
 6 & 03 17 13.6 & $-$41 10 32 & 39.2 & NGC  1291      &  $+$0.0 &      & $-$0.9 &  8.60 & 2 & 10.3 & 24.4 \\
 7 & 03 18 19.9 & $-$66 29 07 & 39.8 & NGC  1313      &  $+$7.0 &      & $+$0.0 &  3.70 & 2 &  0.9 &  9.8 \\
 8 & 03 18 22.2 & $-$66 36 01 & 39.4 & NGC  1313      &  $+$7.0 &      & $+$0.0 &  3.70 & 2 &  6.7 &  9.8 \\
 9 & 03 21 59.8 & $-$37 05 06 & 40.0 & NGC  1316      &  $-$2.0 & $-$1.0 & $-$0.8 & 23.91 & 1 & 77.3 & 83.6 \\
10 & 03 22 40.4 & $-$37 16 41 & 39.7 & NGC  1316      &  $-$2.0 & $-$1.0 & $-$0.8 & 23.91 & 1 & 29.4 & 83.6 \\
11 & 03 22 51.3 & $-$37 09 47 & 39.5 & NGC  1316      &  $-$2.0 & $-$1.0 & $-$0.8 & 23.91 & 1 & 23.0 & 83.6 \\
12 & 03 22 57.3 & $-$37 15 57 & 39.4 & NGC  1316      &  $-$2.0 & $-$1.0 & $-$0.8 & 23.91 & 1 & 32.5 & 83.6 \\
13 & 03 23 14.4 & $-$37 16 51 & 39.4 & NGC  1316      &  $-$2.0 & $-$1.0 & $-$0.8 & 23.91 & 1 & 54.7 & 83.6 \\
14 & 03 23 28.8 & $-$37 10 09 & 39.5 & NGC  1316      &  $-$2.0 & $-$1.0 & $-$0.8 & 23.91 & 1 & 67.3 & 83.6 \\
15 & 03 33 11.9 & $-$36 11 35 & 39.8 & NGC  1365      &  $+$3.0 &  1.8 & $+$0.9 & 22.16   & 1 & 38.5 & 72.3 \\
16 & 03 33 13.0 & $-$36 11 53 & 39.8 & NGC  1365      &  $+$3.0 &  1.8 & $+$0.9 & 22.16   & 1 & 38.5 & 72.3 \\
17 & 03 38 31.8 & $-$35 26 01 & 39.4 & NGC  1399      &  $-$5.0 &      &      & 19.29     & 1 &  6.2 & 38.8 \\
18 & 03 38 32.7 & $-$35 27 01 & 39.7 & NGC  1399      &  $-$5.0 &      &      & 19.29     & 1 &  4.2 & 38.8 \\
19 & 03 38 41.2 & $-$35 31 35 & 40.4 & NGC  1399      &  $-$5.0 &      &      & 19.29     & 1 & 29.4 & 38.8 \\
20 & 03 38 51.7 & $-$35 26 43 & 39.6 & NGC  1399      &  $-$5.0 &      &      & 19.29     & 1 & 26.0 & 38.8 \\
21 & 03 40 12.3 & $-$35 37 36 & 39.9 & NGC  1427A     & $+$10.0 &      & $-$0.2 & 26.99   & 1 &  5.0 & 18.4 \\
22 & 03 45 55.7 & $+$68 04 58 & 39.2 & IC    342      &  $+$6.0 &      & $+$0.5 &  3.90 & 2 &  5.8 & 24.2 \\
23 & 04 15 51.9 & $-$55 49 01 & 39.4 & NGC  1553      &  $-$2.0 & $-$1.0 & $-$1.3 & 14.40 & 1 & 14.2 & 18.7 \\
24 & 04 19 55.6 & $-$54 56 37 & 39.7 & NGC  1566      &  $+$4.0 &  1.0 & $+$0.3 & 19.95   & 1 &  4.4 & 48.3 \\
25 & 04 45 30.9 & $-$59 14 55 & 39.1 & NGC  1672      &  $+$3.0 &  2.0 & $+$0.8 & 18.00   & 1 &  7.6 & 34.6 \\
26 & 04 45 33.6 & $-$59 14 41 & 39.6 & NGC  1672      &  $+$3.0 &  2.0 & $+$0.8 & 18.00   & 1 &  5.9 & 34.6 \\
27 & 04 45 53.2 & $-$59 14 57 & 39.5 & NGC  1672      &  $+$3.0 &  2.0 & $+$0.8 & 18.00   & 1 &  7.4 & 34.6 \\
28 & 05 05 11.1 & $-$37 58 48 & 39.3 & NGC  1792      &  $+$4.0 &      & $+$0.8 & 16.33   & 1 &  3.8 & 24.9 \\
29 & 05 41 43.3 & $+$69 20 46 & 40.5 & NGC  1961      &  $+$5.0 & $-$1.0 & $+$0.6 & 52.40 & 1 & 41.6 & 69.7 \\
30 & 06 15 15.3 & $+$71 02 04 & 40.4 & Mrk 3          &  $-$2.0 &  2.0 & $+$0.8 & 53.31   & 1 & 26.1 & 28.2 \\
31 & 08 19 30.2 & $+$70 42 18 & 40.2 & Holmberg II    & $+$10.0 &      & $-$0.5 &  4.50 & 2 &  2.7 & 10.4 \\
32 & 09 10 19.9 & $+$07 06 00 & 39.2 & NGC  2775      &  $+$2.0 &      & $-$0.1 & 18.05 & 1 & 19.4 & 22.4 \\
33 & 09 10 27.0 & $+$06 59 10 & 39.5 & NGC  2775      &  $+$2.0 &      & $-$0.1 & 18.05 & 1 & 18.6 & 22.4 \\
34 & 09 57 54.4 & $+$69 03 43 & 40.0 & NGC  3031      &  $+$2.0 &  1.8 & $-$0.8 &  3.55 & 3 & 13.0 & 27.8 \\
35 & 10 23 26.0 & $+$19 56 20 & 39.2 & NGC  3226      &  $-$5.0 & $-$1.0 &       & 17.63 & 1 & 12.5 & 16.2 \\
36 & 10 27 37.6 & $-$43 53 08 & 39.8 & NGC  3256      & $+$99.0 &      &  $+$1.7 & 37.08 & 1 & 30.0 & 41.0 \\
37 & 10 37 39.6 & $-$27 05 23 & 40.9 & IC   2597      &  $-$4.0 &      &         & 40.09 & 1 & 21.7 & 30.0 \\
38 & 10 38 43.4 & $+$53 31 06 & 39.5 & NGC  3310      &  $+$4.0 &      &  $+$0.8 & 18.70 & 2 &  5.7 & 16.8 \\
39 & 11 20 15.5 & $+$13 35 25 & 39.5 & NGC  3628      &  $+$3.0 &      &  $+$0.8 &  7.70 & 2 &  0.5 & 33.1 \\
40 & 11 50 57.9 & $-$28 44 02 & 39.9 & NGC  3923      &  $-$5.0 &      &      & 22.24 & 1 & 28.8 & 38.1 \\
41 & 12 01 40.4 & $+$62 19 58 & 39.6 & UGC  7009      & $+$10.0 &      &      & 14.93 & 1 &  2.4 &  7.0 \\
42 & 12 05 32.6 & $+$50 32 46 & 39.5 & NGC  4088      &  $+$4.0 &      &  $+$0.8 & 17.00 & 2 &  2.5 & 28.5 \\
43 & 12 10 07.9 & $+$39 23 12 & 39.3 & NGC  4151      &  $+$2.0 &  1.5 &         & 20.30 & 2 & 29.6 & 37.3 \\
44 & 12 10 22.6 & $+$39 23 16 & 39.0 & NGC  4151      &  $+$2.0 &  1.5 &         & 20.30 & 2 & 13.8 & 37.3 \\
45 & 12 15 15.7 & $+$33 10 20 & 39.0 & NGC  4203      &  $-$3.0 & $-$1.0 & $-$0.4 & 14.48 & 1 & 11.2 & 14.3 \\
46 & 12 18 56.1 & $+$14 24 18 & 40.5 & NGC  4254      &  $+$5.0 &      &  $+$0.5 & 32.09 & 1 & 16.9 & 50.1 \\
47 & 12 22 43.0 & $+$15 44 00 & 39.5 & NGC  4321      &  $+$4.0 &      &  $+$0.3 & 21.15 & 1 & 37.7 & 45.6 \\
48 & 12 22 48.4 & $+$15 43 09 & 39.2 & NGC  4321      &  $+$4.0 &      &  $+$0.3 & 21.15 & 1 & 39.6 & 45.6 \\
49 & 12 23 08.6 & $+$15 51 23 & 39.3 & NGC  4321      &  $+$4.0 &      &  $+$0.3 & 21.15 & 1 & 23.3 & 45.6 \\
50 & 12 25 05.1 & $+$12 53 06 & 40.0 & NGC  4374      &  $-$5.0 &      & $-$1.2 & 16.80 & 2 &  1.8 & 31.6 \\
51 & 12 25 15.4 & $+$12 56 02 & 39.6 & NGC  4374      &  $-$5.0 &      & $-$1.2 & 16.80 & 2 & 19.5 & 31.6 \\
52 & 12 25 16.8 & $-$39 43 26 & 40.0 & NGC  4373      &  $-$2.9 &      &      & 45.28 & 1 & 29.1 & 44.6 \\
53 & 12 26 01.8 & $+$33 31 34 & 39.0 & NGC  4395      &  $+$9.0 &  1.8 & $-$0.4   &  3.60 & 2 &  2.9 & 13.8 \\
54 & 12 27 57.5 & $+$13 02 30 & 39.9 & NGC  4438      &  $+$0.0 & $-$1.0 & $-$0.1 & 16.80 & 2 & 17.0 & 41.6 \\
55 & 12 29 13.1 & $+$07 57 40 & 39.1 & NGC  4472      &  $-$5.0 &  2.0 &          & 16.80 & 2 & 42.0 & 50.0 \\
56 & 12 29 22.3 & $+$07 53 31 & 39.4 & NGC  4472      &  $-$5.0 &  2.0 &          & 16.80 & 2 & 43.0 & 50.0 \\
57 & 12 29 23.9 & $+$07 54 00 & 39.8 & NGC  4472      &  $-$5.0 &  2.0 &          & 16.80 & 2 & 40.0 & 50.0 \\
58 & 12 29 27.9 & $+$08 06 34 & 39.4 & NGC  4472      &  $-$5.0 &  2.0 &          & 16.80 & 2 & 39.3 & 50.0 \\
59 & 12 29 35.1 & $+$08 09 37 & 39.3 & NGC  4472      &  $-$5.0 &  2.0 &          & 16.80 & 2 & 49.1 & 50.0 \\
60 & 12 29 39.9 & $+$07 53 32 & 39.9 & NGC  4472      &  $-$5.0 &  2.0 &          & 16.80 & 2 & 32.4 & 50.0 \\
61 & 12 30 06.7 & $+$08 02 05 & 39.2 & NGC  4472      &  $-$5.0 &  2.0 &          & 16.80 & 2 & 26.5 & 50.0 \\
62 & 12 30 30.8 & $+$41 41 45 & 39.3 & NGC  4485      & $+$10.0 &      &          &  9.30 & 2 &  0.8 &  6.2 \\
63 & 12 35 19.2 & $+$12 33 17 & 39.5 & NGC  4552      &  $-$5.0 &      &          & 16.80 & 2 & 24.7 & 25.1 \\
64 & 12 35 29.4 & $+$12 31 10 & 39.3 & NGC  4552      &  $-$5.0 &      &          & 16.80 & 2 & 16.7 & 25.1 \\
65 & 12 35 51.7 & $+$27 56 04 & 39.9 & NGC  4559      &  $+$6.0 &      & $-$0.0   &  9.70 & 2 &  5.8 & 30.2 \\
66 & 12 35 58.7 & $+$27 57 42 & 39.7 & NGC  4559      &  $+$6.0 &      & $-$0.0 &  9.70 & 2 & 0.6 & 30.2 \\
67 & 12 36 17.4 & $+$25 58 56 & 40.1 & NGC  4565      &  $+$3.0 &  1.9 &  $+$0.0 & 16.36 & 1 & 3.5 & 75.4 \\
68 & 12 41 55.5 & $+$32 32 14 & 39.1 & NGC  4631      &  $+$7.0 &      & $+$0.6 &  6.90 & 2 & 5.2 & 31.1 \\
69 & 12 43 36.8 & $+$11 30 06 & 39.4 & NGC  4649      &  $-$5.0 &      &      & 14.85 & 1 & 12.9 & 32.0 \\
70 & 12 44 07.2 & $+$11 35 25 & 39.3 & NGC  4649      &  $-$5.0 &      &      & 14.85 & 1 & 30.4 & 32.0 \\
71 & 12 44 09.2 & $+$11 33 36 & 39.7 & NGC  4649      &  $-$5.0 &      &      & 14.85 & 1 & 30.7 & 32.0 \\
72 & 12 59 00.9 & $+$34 50 42 & 39.8 & NGC  4861      &  $+$9.0 &      &           & 17.80 & 2 &  5.8 & 20.6 \\
73 & 12 59 02.0 & $+$34 51 11 & 40.3 & NGC  4861      &  $+$9.0 &      &           & 17.80 & 2 &  3.0 & 20.6 \\
74 & 13 15 19.6 & $+$42 02 58 & 39.6 & NGC  5055      &  $+$4.0 & $-$1.0 &  $+$0.4 &  7.20 & 2 & 11.7 & 26.4 \\
75 & 13 25 07.4 & $-$43 04 06 & 39.3 & NGC  5128      &  $-$2.0 &  2.0 &  $+$0.3   &  4.90 & 2 &  7.2 & 36.6 \\
76 & 13 25 19.9 & $-$43 03 13 & 39.9 & NGC  5128      &  $-$2.0 &  2.0 &  $+$0.3   &  4.90 & 2 &  4.0 & 36.6 \\
77 & 13 29 38.7 & $+$58 25 06 & 39.5 & NGC  5204      &  $+$9.0 &      &  $+$0.0   &  4.80 & 2 &  0.4 &  7.0 \\
78 & 13 29 40.2 & $+$47 12 41 & 39.0 & NGC  5194      &  $+$4.0 & 2.0 &  $+$0.3    &  8.40 & 2 &  5.8 & 25.1 \\
79 & 13 29 43.6 & $+$47 11 36 & 39.1 & NGC  5194      &  $+$4.0 & 2.0 &  $+$0.3    &  8.40 & 2 &  4.1 & 25.1 \\
80 & 13 30 01.3 & $+$47 13 42 & 39.4 & NGC  5194      &  $+$4.0 &  2.0 &  $+$0.3   &  8.40 & 2 &  5.7 & 25.1 \\
81 & 13 30 07.2 & $+$47 11 02 & 39.3 & NGC  5194      &  $+$4.0 &  2.0 &  $+$0.3   &  8.40 & 2 &  6.1 & 25.1 \\
82 & 13 37 20.1 & $-$29 53 46 & 39.2 & NGC  5236      &  $+$5.0 &      &  $+$0.3   &  4.70 & 2 &  6.3 & 17.6 \\
83 & 14 04 14.3 & $+$54 26 05 & 39.1 & NGC  5457      &  $+$6.0 &      &  $+$0.3   &  5.40 & 2 & 16.3 & 45.3 \\
84 & 14 53 44.7 & $+$03 33 30 & 39.4 & NGC  5775      &  $+$5.0 &      &  $+$1.1 & 22.41 & 1 & 21.8 & 27.2 \\
85 & 20 35 00.3 & $+$60 09 05 & 39.0 & NGC  6946      &  $+$6.0 &      &  $+$0.6 &  5.50 & 2 &  1.7 & 18.4 \\
86 & 22 35 48.3 & $-$26 01 26 & 39.4 & NGC  7314      &  $+$4.0 &  1.9 &  $+$0.3 & 18.96 & 1 & 9.4 & 25.2 \\
87 & 23 18 55.9 & $-$42 13 53 & 39.8 & NGC  7590      &  $+$4.0 &  2.0 &      & 21.28 & 1 & 2.7 & 16.7 \\
\enddata

\tablenotetext{\ }{
\noindent
Notes on table columns:
(1) IXO catalog number;
(2,3) X-ray position from ROSAT HRI data;
(4) log of maximum 2$-$10 keV X-ray luminosity, calculated from the ROSAT HRI 
count rate assuming a power-law spectrum with $\Gamma=$1.7 and the Galactic hydrogen
column from Table 2.
We note that while a soft X-ray luminosity (e.g., 0.2$-$2.4 keV)
is better matched to the ROSAT HRI instrument than a hard 2$-$10 keV luminosity,
the latter is better matched to current missions, 
such as {\it Chandra} and {\it XMM}.
For $\Gamma=$1.7, the 0.2$-$2.4 keV unabsorbed luminosity
is a factor of 0.89 times the 2$-$10 keV luminosity listed here;
(5) Galaxy Name;
(6) Host galaxy type from RC3 (T-type, negative for E galaxies, positive for
S galaxies);
(7) AGN type (-1 means normal galaxy);
(8) log of FIR to B-band flux of galaxy.  FIR flux determined from IRAS
    measurements using the method of Fullmer \& Lonsdale (1989).
    B-band fluxes were extracted from NED;
(9,10) Distance to galaxy in Mpc.  References: 
[1] H$_0 =$ 75 km~s$^{-1}$~Mpc$^{-1}$ using recessional velocities listed in RC3,
[2] Nearby Galaxies Catalog, Tully (1988),
[3] Paturel et al. (2002);
(11) radius from nucleus, which was assumed to be at the RC3 optical position;
(12) maximum radius allowed for inclusion to this galaxy, equal to the 
     25 mag~asec$^{-2}$ diameter from RC3.
}

\end{deluxetable}

\clearpage

\begin{deluxetable}{rlcrrr}
\tablecaption{ROSAT HRI Detections of IXOs\label{tab2}}
\tablewidth{0pt}
\tablehead{
\colhead{IXO} & \colhead{Dataset}   & \colhead{Count Rate}& 
\colhead{Significance} & \colhead{N$_H$} &
\colhead{L$_X$} \\
\colhead{} & \colhead{} & \multicolumn{1}{c}{(10$^{-3}$ s$^{-1}$)} & \colhead{} &
\colhead{($log$ cm$^{-2}$)} &
\colhead{(10$^{39}$ erg~s$^{-1}$)} \\
\colhead{(1)} & \colhead{(2)} & \colhead{(3)} & \colhead{(4)} & 
\colhead{(5)} & \colhead{(6)} \\
}
\startdata
 1   &   rh600472n00\tablenotemark{*} & 0.50$\pm$0.02 &   3.9   &  20.2  &   1.38 \\ % PGC 06983  orignh=  20.2 
 2   &   rh600472n00\tablenotemark{*}   & 1.77$\pm$0.08 &   13.3   &  20.2  &   4.89 \\ % PGC 06983  orignh=  20.2 
 3   &   rh500501n00\tablenotemark{*}   & 2.13$\pm$0.13 &   10.7   &  20.9  &   1.52 \\ % PGC 09031  orignh=  20.9 
     &   rh600690n00   & 1.41$\pm$0.05 &   12.8   &        &   1.01 \\ % PGC 00000  orignh=  99.9 
 4   &   rh600469a01\tablenotemark{*}   & 6.04$\pm$0.36 &   17.2   &  20.5  &   12.51 \\ % PGC 10122  orignh=  20.5 
 5   &   rh600999n00\tablenotemark{*}   & 0.65$\pm$0.05 &   4.2   &  20.6  &   1.13 \\ % PGC 10329  orignh=  20.6 
 6   &   rh600828n00\tablenotemark{*}   & 3.81$\pm$0.30 &   14.6   &  20.3  &   1.59 \\ % PGC 12209  orignh=  20.3 
 7   &   rh400065n00   & 68.50$\pm$3.29 &   83.6   &  20.6  &   6.11 \\ % PGC 12286  orignh=  20.6 
     &   rh500403n00   & 63.20$\pm$2.02 &   149.5   &        &   5.64 \\ % PGC 00000  orignh=  99.9 
     &   rh500403a01\tablenotemark{*}   & 29.70$\pm$0.85 &   98.2   &        &   2.65 \\ % PGC 00000  orignh=  99.9 
     &   rh500404n00   & 64.80$\pm$1.43 &   187.1   &        &   5.78 \\ % PGC 00000  orignh=  99.9 
     &   rh500404a01   & 36.90$\pm$1.26 &   88.3   &        &   3.29 \\ % PGC 00000  orignh=  99.9 
     &   rh500492n00   & 62.90$\pm$1.53 &   182.1   &        &   5.61 \\ % PGC 00000  orignh=  99.9 
     &   rh500550n00   & 28.20$\pm$0.93 &   84.5   &        &   2.52 \\ % PGC 00000  orignh=  99.9 
     &   rh600505n00   & 49.20$\pm$1.36 &   161.8   &        &   4.39 \\ % PGC 00000  orignh=  99.9 
     &   rh600505a01   & 62.20$\pm$1.62 &   161.8   &        &   5.55 \\ % PGC 00000  orignh=  99.9 
 8   &   rh400065n00   & 18.70$\pm$1.62 &   43.9   &  20.6  &   1.67 \\ % PGC 12286  orignh=  20.6 
     &   rh500403n00   & 25.30$\pm$1.25 &   73.6   &        &   2.26 \\ % PGC 00000  orignh=  99.9 
     &   rh500403a01\tablenotemark{*}   & 11.50$\pm$0.50 &   49.0   &        &   1.03 \\ % PGC 00000  orignh=  99.9 
     &   rh500404n00   & 17.50$\pm$0.69 &   65.2   &        &   1.56 \\ % PGC 00000  orignh=  99.9 
     &   rh500404a01   & 15.00$\pm$0.74 &   42.4   &        &   1.34 \\ % PGC 00000  orignh=  99.9 
     &   rh500492n00   & 16.40$\pm$0.74 &   76.6   &        &   1.46 \\ % PGC 00000  orignh=  99.9 
     &   rh500550n00   & 15.70$\pm$0.67 &   58.9   &        &   1.40 \\ % PGC 00000  orignh=  99.9 
     &   rh600505n00   & 8.45$\pm$0.46 &   33.2   &        &   0.75 \\ % PGC 00000  orignh=  99.9 
 9   &   rh600255a01\tablenotemark{*}   & 3.30$\pm$0.23 &   9.6   &  20.3  &   10.39 \\ % PGC 12651  orignh=  20.3 
10   &   rh600255a01\tablenotemark{*}   & 1.51$\pm$0.11 &   9.7   &  20.3  &   4.76 \\ % PGC 12651  orignh=  20.3 
11   &   rh600255n00                    & 0.97$\pm$0.11 &   3.7   &  20.3  &   3.05 \\ % PGC 00000  orignh=  99.9 
     &   rh600255a01\tablenotemark{*}   & 1.01$\pm$0.08 &   5.5   &        &   3.18 \\ % PGC 12651  orignh=  20.3 
12   &   rh600255a01\tablenotemark{*}   & 0.75$\pm$0.06 &   3.1   &  20.3  &   2.36 \\ % PGC 12651  orignh=  20.3 
13   &   rh600255a01\tablenotemark{*}   & 0.80$\pm$0.06 &   3.9   &  20.3  &   2.53 \\ % PGC 12651  orignh=  20.3 
14   &   rh600255a01\tablenotemark{*}   & 1.12$\pm$0.09 &   3.3   &  20.3  &   3.53 \\ % PGC 12651  orignh=  20.3 
15   &   rh701297n00                    & 2.41$\pm$0.30 &   5.4   &  20.1  &   6.11 \\ % PGC 13179  orignh=  20.1 
     &   rh701297a02\tablenotemark{*}   & 2.64$\pm$0.32 &   4.2   &        &   6.69 \\ % PGC 00000  orignh=  99.9 
16   &   rh701297n00                     & 2.56$\pm$0.30 &   6.6   & 20.1       &   6.48 \\ % PGC 00000  orignh=  99.9 
     &   rh701297a02\tablenotemark{*}   & 2.14$\pm$0.28 &   6.7   &    &   5.42 \\ % PGC 13179  orignh=  20.1 
17   &   rh600940n00\tablenotemark{*}   & 1.08$\pm$0.03 &   3.1   &  20.1  &   2.06 \\ % PGC 13418  orignh=  20.1 
     &   rh600831a01                    & 1.21$\pm$0.03 &   5.0   &        &   2.31 \\ % PGC 00000  orignh=  99.9 
18   &   rh600831n00   & 2.53$\pm$0.06 &   3.6   &  20.1  &   4.82 \\ % PGC 00000  orignh=  99.9 
     &   rh600831a01   & 2.34$\pm$0.06 &   3.8   &        &   4.46 \\ % PGC 00000  orignh=  99.9 
     &   rh600940n00\tablenotemark{*}   & 2.76$\pm$0.07 &   3.5   &        &   5.26 \\ % PGC 13418  orignh=  20.1 
19   &   rh600220n00   & 12.00$\pm$0.96 &   26.8   & 20.1  &   22.88 \\ % PGC 00000  orignh=  99.9 
     &   rh600256n00   & 9.94$\pm$0.88 &   22.3   &        &   18.95 \\ % PGC 00000  orignh=  99.9 
     &   rh600831n00   & 3.86$\pm$0.14 &   22.6   &        &   7.36 \\ % PGC 13418  orignh=  20.1 
     &   rh600831a01   & 7.75$\pm$0.21 &   53.7   &        &   14.78 \\ % PGC 00000  orignh=  99.9 
     &   rh600940n00\tablenotemark{*}   & 3.70$\pm$0.13 &   25.7   &        &   7.05 \\ % PGC 00000  orignh=  99.9 
20   &   rh600940n00\tablenotemark{*}   & 1.67$\pm$0.07 &   7.5   &   20.1     &   3.18 \\ % PGC 00000  orignh=  99.9 
     &   rh600831a01                    & 2.33$\pm$0.08 &   14.2   &       &   4.44 \\ % PGC 13418  orignh=  20.1 
21   &   rh600940n00\tablenotemark{*}   & 1.96$\pm$0.08 &   3.5   &  20.1  &   7.34 \\ % PGC 13500  orignh=  20.1 
22   &   rh600022n00\tablenotemark{*}   & 7.38$\pm$0.47 &   26.7   &  21.5  &   1.52 \\ % PGC 13826  orignh=  21.5 
23   &   rh600479n00\tablenotemark{*}   & 2.54$\pm$0.14 &   16.5   &  20.2  &   2.76 \\ % PGC 14765  orignh=  20.2 
     &   rh600490a01   & 1.88$\pm$0.12 &   11.5   &        &   2.04 \\ % PGC 00000  orignh=  99.9 
24   &   rh600860n00   & 2.05$\pm$0.15 &   4.4   &  20.1  &   4.18 \\ % PGC 00000  orignh=  99.9 
     &   rh600962n00\tablenotemark{*}   & 2.39$\pm$0.10 &   8.4   &        &   4.87 \\ % PGC 14897  orignh=  20.1 
25   &   rh601011n00\tablenotemark{*}   & 0.63$\pm$0.04 &   3.0   &  20.4  &   1.17 \\ % PGC 15941  orignh=  20.4 
26   &   rh601011n00\tablenotemark{*}   & 1.92$\pm$0.12 &   7.0   &  20.4  &   3.57 \\ % PGC 15941  orignh=  20.4 
     &   rh701022n00   & 0.82$\pm$0.06 &   4.5   &        &   1.53 \\ % PGC 00000  orignh=  99.9 
27   &   rh601011n00\tablenotemark{*}   & 1.63$\pm$0.10 &   8.0   &  20.4  &   3.03 \\ % PGC 00000  orignh=  99.9 
     &   rh701022n00   & 1.86$\pm$0.14 &   6.8   &        &   3.46 \\ % PGC 15941  orignh=  20.4 
28   &   rh600433n00\tablenotemark{*}   & 1.34$\pm$0.08 &   6.3   &  20.4  &   2.11 \\ % PGC 16709  orignh=  20.4 
29   &   rh600499n00\tablenotemark{*}   & 1.62$\pm$0.07 &   14.3   &  20.9  &   35.23 \\ % PGC 17625  orignh=  20.9 
30   &   rh700339n00\tablenotemark{*}   & 1.20$\pm$0.12 &   4.9   &  20.9  &   27.10 \\ % PGC 18722  orignh=  20.9 
31   &   rh600745n00\tablenotemark{*}   & 118.00$\pm$3.63 &   136.5   &  20.5  &   15.10 \\ % PGC 23324  orignh=  20.5 
32   &   rh600826a01\tablenotemark{*}   & 0.77$\pm$0.05 &   3.8   &  20.6  &   1.66 \\ % PGC 25861  orignh=  20.6 
33   &   rh600826a01\tablenotemark{*}   & 1.35$\pm$0.08 &   7.8   &  20.6  &   2.90 \\ % PGC 25861  orignh=  20.6 
34   &   rh600247n00   & 45.40$\pm$1.27 &   71.3   &  20.6  &   3.79 \\ % PGC 00000  orignh=  99.9 
     &   rh600247a01   & 65.10$\pm$1.70 &   84.4   &        &   5.44 \\ % PGC 00000  orignh=  99.9 
     &   rh600739n00   & 35.50$\pm$1.27 &   49.0   &        &   2.96 \\ % PGC 00000  orignh=  99.9 
     &   rh600740n00   & 57.40$\pm$1.66 &   65.7   &        &   4.79 \\ % PGC 00000  orignh=  99.9 
     &   rh600881n00   & 47.50$\pm$1.74 &   60.3   &        &   3.97 \\ % PGC 00000  orignh=  99.9 
     &   rh600882n00   & 61.60$\pm$1.76 &   67.4   &        &   5.14 \\ % PGC 00000  orignh=  99.9 
     &   rh600882a01\tablenotemark{*}   & 109.00$\pm$4.54 &   56.3   &        &   9.10 \\ % PGC 28630  orignh=  20.6 
     &   rh601001n00   & 112.00$\pm$2.39 &   126.4   &        &   9.36 \\ % PGC 00000  orignh=  99.9 
     &   rh601002n00   & 39.10$\pm$1.33 &   48.6   &        &   3.27 \\ % PGC 00000  orignh=  99.9 
     &   rh601095n00   & 41.30$\pm$1.63 &   31.4   &        &   3.45 \\ % PGC 00000  orignh=  99.9 
35   &   rh701299n00\tablenotemark{*}   & 0.88$\pm$0.06 &   4.7   &  20.3  &   1.55 \\ % PGC 30440  orignh=  20.3 
36   &   rh701983n00\tablenotemark{*}   & 0.59$\pm$0.03 &   4.9   &  21.0  &   6.73 \\ % PGC 30785  orignh=  21.0 
37   &   rh800741n00\tablenotemark{*}   & 7.07$\pm$0.35 &   33.2   &  20.7  &   79.12 \\ % PGC 31586  orignh=  20.7 
38   &   rh600685a01\tablenotemark{*}   & 1.89$\pm$0.10 &   8.7   &  20.1  &   3.27 \\ % PGC 31650  orignh=  20.1 
39   &   rh700009n00\tablenotemark{*}   & 10.40$\pm$0.72 &   42.6   &  20.3  &   3.52 \\ % PGC 34697  orignh=  20.3 
40   &   rh600679n00\tablenotemark{*}   & 1.78$\pm$0.10 &   11.2   &  20.8  &   6.45 \\ % PGC 37061  orignh=  20.8 
     &   rh600679a01                    & 2.15$\pm$0.16 &   10.4   &        &   7.79 \\ % PGC 00000  orignh=  99.9 
41   &   rh500499n00\tablenotemark{*}   & 3.08$\pm$0.17 &   8.9   &  20.3  &   3.83 \\ % PGC 37951  orignh=  20.3 
42   &   rh500391n00\tablenotemark{*}   & 1.91$\pm$0.18 &   4.6   &  20.3  &   3.08 \\ % PGC 38302  orignh=  20.3 
43   &   rh701707n00\tablenotemark{*}   & 0.88$\pm$0.03 &   9.2   &  20.3  &   2.02 \\ % PGC 38739  orignh=  20.3 
44   &   rh701707n00\tablenotemark{*}   & 0.47$\pm$0.02 &   4.9   &  20.3  &   1.09 \\ % PGC 38739  orignh=  20.3 
45   &   rh600221n00\tablenotemark{*}   & 0.98$\pm$0.08 &   4.1   &  20.1  &   1.02 \\ % PGC 39158  orignh=  20.1 
46   &   rh600972n00\tablenotemark{*}   & 5.08$\pm$0.32 &   17.9   &  20.4  &   31.32 \\ % PGC 39578  orignh=  20.4 
47   &   rh500542n00   & 1.03$\pm$0.08 &   4.9   &  20.4  &   2.68 \\ % PGC 40153  orignh=  20.4 
     &   rh600731n00\tablenotemark{*}   & 1.11$\pm$0.07 &   7.1   &        &   2.89 \\ % PGC 00000  orignh=  99.9 
48   &   rh600731n00\tablenotemark{*}   & 0.62$\pm$0.04 &   3.4   &  20.4  &   1.62 \\ % PGC 40153  orignh=  20.4 
49   &   rh500542n00\tablenotemark{*}   & 0.77$\pm$0.06 &   3.5   &  20.4  &   2.00 \\ % PGC 40153  orignh=  20.4 
50   &   rh600493n00\tablenotemark{*} \\
     &   rh700192n00   & 6.56$\pm$0.38 &   3.1   &  20.4  &   10.97 \\ % PGC 40455  orignh=  20.4 
51   &   rh600493n00\tablenotemark{*}   & 2.13$\pm$0.15 &   8.6   &  20.4  &   3.56 \\ % PGC 40455  orignh=  20.4 
52   &   rh600976n00\tablenotemark{*}   & 0.65$\pm$0.04 &   3.6   &  20.8  &   9.94 \\ % PGC 40498  orignh=  20.8 
53   &   rh702725n00\tablenotemark{*}   & 15.20$\pm$1.00 &   48.3   &  20.1  &   1.01 \\ % PGC 40596  orignh=  20.1 
54   &   rh600608n00\tablenotemark{*}   & 5.13$\pm$0.33 &   18.5   &  20.4  &   8.62 \\ % PGC 40914  orignh=  20.4 
     &   rh701657n00   & 3.95$\pm$0.52 &   6.1   &        &   6.64 \\ % PGC 00000  orignh=  99.9 
55   &   rh600216a01\tablenotemark{*}   & 0.88$\pm$0.07 &   3.5   &  20.2  &   1.34 \\ % PGC 41220  orignh=  20.2 
56   &   rh600216a01\tablenotemark{*}   & 1.69$\pm$0.12 &   5.8   &  20.2  &   2.56 \\ % PGC 41220  orignh=  20.2 
57   &   rh600216n00                    & 4.29$\pm$0.47 &   7.4   &  20.2      &   6.48 \\ % PGC 00000  orignh=  99.9 
     &   rh600216a01\tablenotemark{*}   & 4.52$\pm$0.26 &   16.7   &        &   6.83 \\ % PGC 41220  orignh=  20.2 
58   &   rh600216a01\tablenotemark{*}   & 1.70$\pm$0.12 &   6.3   &  20.2  &   2.57 \\ % PGC 41220  orignh=  20.2 
59   &   rh600216a01\tablenotemark{*}   & 1.47$\pm$0.12 &   4.1   &  20.2  &   2.22 \\ % PGC 41220  orignh=  20.2 
60   &   rh600216n00                    & 5.63$\pm$0.65 &   11.3   &  20.2  &   8.52 \\ % PGC 00000  orignh=  99.9 
     &   rh600216a01\tablenotemark{*}   & 5.13$\pm$0.28 &   21.3   &        &   7.76 \\ % PGC 41220  orignh=  20.2 
61   &   rh600216a01\tablenotemark{*}   & 1.13$\pm$0.08 &   3.0   &  20.2  &   1.71 \\ % PGC 41220  orignh=  20.2 
62   &   rh600697n00   & 4.45$\pm$0.51 &   4.4   &  20.3  &   2.10 \\ % PGC 00000  orignh=  99.9 
     &   rh600855n00\tablenotemark{*}   & 2.52$\pm$0.15 &   11.7   &        &   1.19 \\ % PGC 41326  orignh=  20.3 
     &   rh600855a01   & 1.87$\pm$0.14 &   9.8   &        &   0.88 \\ % PGC 00000  orignh=  99.9 

63   &   rh600491n00\tablenotemark{*} \\
     &   rh600491a01   & 2.03$\pm$0.14 &   8.5   &  20.4  &   3.38 \\ % PGC 41968  orignh=  20.4 
64   &   rh600491n00\tablenotemark{*}   & 1.10$\pm$0.10 &   3.0   &  20.4  &   1.83 \\ % PGC 41968  orignh=  20.4 
65   &   rh600861n00\tablenotemark{*}   & 15.20$\pm$0.63 & 91.8 & 20.2 & 7.49 \\
66   &   rh600861n00\tablenotemark{*}   & 9.18$\pm$0.31 &   48.9   &  20.2  &   4.52 \\ % PGC 42002  orignh=  20.2 
67   &   rh600691n00\tablenotemark{*}   & 8.63$\pm$0.97 &   16.0   &  20.1  &   11.76 \\ % PGC 42038  orignh=  20.1 
68   &   rh600193a00   & 2.40$\pm$0.30 &   7.0   &  20.1  &   0.58 \\ % PGC 42620  orignh=  20.1 
     &   rh600193a01   & 2.70$\pm$0.22 &   8.7   &        &   0.65 \\ % PGC 00000  orignh=  99.9 
     &   rh600193a02   & 5.36$\pm$0.47 &   19.2   &        &   1.29 \\ % PGC 00000  orignh=  99.9 
     &   rh600934n00\tablenotemark{*}   & 3.74$\pm$0.24 &   16.1   &        &   0.90 \\ % PGC 00000  orignh=  99.9 
69   &   rh600480a01   & 2.17$\pm$0.21 &   4.4   &  20.3  &   2.73 \\ % PGC 42831  orignh=  20.3 
     &   rh600620a01\tablenotemark{*}   & 2.21$\pm$0.16 &   7.4   &        &   2.78 \\ % PGC 00000  orignh=  99.9 
70   &   rh600620a01\tablenotemark{*}   & 1.43$\pm$0.12 &   4.4   &  20.3  &   1.80 \\ % PGC 42831  orignh=  20.3 
71   &   rh600480a01   & 2.56$\pm$0.29 &   9.4   &  20.3  &   3.22 \\ % PGC 00000  orignh=  99.9 
     &   rh600620a01\tablenotemark{*}   & 4.35$\pm$0.31 &   14.1   &        &   5.47 \\ % PGC 42831  orignh=  20.3 
72   &   rh600273n00                    & 3.58$\pm$0.42 &   8.3   &  20.1  &   5.67 \\ % PGC 00000  orignh=  99.9 
     &   rh600273a01\tablenotemark{*}   & 2.28$\pm$0.28 &   8.8   &        &   3.61 \\ % PGC 44536  orignh=  20.1 
73   &   rh600273n00                    & 2.18$\pm$0.25 &   8.6   & 20.1  &   3.46 \\ % PGC 00000  orignh=  99.9 
     &   rh600273a01\tablenotemark{*}   & 11.50$\pm$0.90 &   39.3   &        &   18.23 \\ % PGC 44536  orignh=  20.1 
74   &   rh600742n00\tablenotemark{*}   & 13.80$\pm$0.87 &   41.1   &  20.1  &   3.65 \\ % PGC 46153  orignh=  20.1 
     &   rh601110n00   & 7.80$\pm$0.64 &   23.6   &        &   2.06 \\ % PGC 00000  orignh=  99.9 
75   &   rh150004n00                    & 5.54$\pm$0.36 &   20.0   &  20.9      &   1.06 \\ % PGC 00000  orignh=  99.9 
     &   rh701924n00   & 11.00$\pm$1.17 &   18.3   &        &   2.11 \\ % PGC 00000  orignh=  99.9 
     &   rh701925n00\tablenotemark{*}   & 9.32$\pm$0.95 &   15.3   &   &   1.79 \\ % PGC 46957  orignh=  20.9 
     &   rh701926n00   & 5.32$\pm$0.63 &   7.3   &        &   1.02 \\ % PGC 00000  orignh=  99.9 
     &   rh701928n00   & 7.95$\pm$1.04 &   8.3   &        &   1.53 \\ % PGC 00000  orignh=  99.9 
     &   rh704206n00   & 5.49$\pm$0.35 &   18.3   &        &   1.05 \\ % PGC 00000  orignh=  99.9 
76   &   rh701924n00                    & 32.20$\pm$2.20 &   45.9   &  20.9  &   6.20 \\ % PGC 00000  orignh=  99.9 
     &   rh701925n00\tablenotemark{*}   & 37.00$\pm$2.40 &   60.1   &        &   7.12 \\ % PGC 46957  orignh=  20.9 
     &   rh701926n00   & 32.40$\pm$2.10 &   48.0   &        &   6.23 \\ % PGC 00000  orignh=  99.9 
     &   rh701927n00   & 24.60$\pm$2.12 &   31.8   &        &   4.73 \\ % PGC 00000  orignh=  99.9 
     &   rh701928n00   & 33.70$\pm$2.55 &   34.6   &        &   6.48 \\ % PGC 00000  orignh=  99.9 
77   &   rh702723n00                    & 23.10$\pm$1.06 &   61.2   &  20.1  &   2.75 \\ % PGC 00000  orignh=  99.9 
     &   rh702723a01\tablenotemark{*}   & 28.30$\pm$1.28 &   61.8   &        &   3.37 \\ % PGC 47368  orignh=  20.1 
78   &   rh600062a01\tablenotemark{*}   & 2.73$\pm$0.30 &   5.1   &  20.2  &   1.02 \\ % PGC 47413  orignh=  20.2 
     &   rh600601n00                    & 2.55$\pm$0.15 &   9.6   &        &   0.95 \\ % PGC 00000  orignh=  99.9 
79   &   rh600062a01\tablenotemark{*}   & 2.70$\pm$0.27 &   6.3   &  20.2  &   1.00 \\ % PGC 47413  orignh=  20.2 
     &   rh600601n00                    & 0.90$\pm$0.05 &   4.6   &        &   0.33 \\ % PGC 00000  orignh=  99.9 
     &   rh601115n00                    & 3.02$\pm$0.32 &   7.0   &        &   1.13 \\ % PGC 00000  orignh=  99.9 
80   &      wh600062                    & 5.50$\pm$0.51 &   11.2  &  20.2  &   2.06 \\ % PGC 00000  orignh=  99.9 
     &   rh600062a00                    & 5.37$\pm$0.49 &   11.0  &        &   2.01 \\ % PGC 00000  orignh=  99.9 
     &   rh600062a01\tablenotemark{*}   & 6.42$\pm$0.55 &   11.5  &        &   2.39 \\ % PGC 00000  orignh=  99.9 
     &   rh600601n00                    & 4.27$\pm$0.20 &   17.2  &        &   1.59 \\ % PGC 47404  orignh=  20.2 
81   &      wh600062                    & 3.91$\pm$0.42 &   9.3   &  20.2  &   1.46 \\ % PGC 47404  orignh=  20.2 
     &   rh600062a00                    & 3.90$\pm$0.42 &   9.4   &        &   1.46 \\ % PGC 00000  orignh=  99.9 
     &   rh600062a01\tablenotemark{*}   & 4.86$\pm$0.47 &   12.4  &        &   1.82 \\ % PGC 00000  orignh=  99.9 
     &   rh600062a03                    & 3.49$\pm$0.36 &   9.6   &        &   1.31 \\ % PGC 00000  orignh=  99.9 
     &   rh600601n00                    & 5.82$\pm$0.28 &   27.2  &        &   2.18 \\ % PGC 00000  orignh=  99.9 
82   &   rh600024n00                    & 5.92$\pm$0.35 &   22.9   & 20.6  &   0.85 \\ % PGC 00000  orignh=  99.9 
     &   rh600024a01\tablenotemark{*}   & 10.30$\pm$0.51 &   40.7   &        &   1.48 \\ % PGC 48082  orignh=  20.6 
83   &   rh600092n00   & 5.65$\pm$0.41 &   13.6   &  20.1  &   0.82 \\ % PGC 00000  orignh=  99.9 
     &   rh600383n00\tablenotemark{*}   & 8.37$\pm$0.41 &   22.2   &        &   1.22 \\ % PGC 50063  orignh=  20.1 
     &   rh600820n00   & 8.52$\pm$0.23 &   38.1   &        &   1.24 \\ % PGC 00000  orignh=  99.9 
     &   rh600820a01   & 4.57$\pm$0.17 &   19.1   &        &   0.67 \\ % PGC 00000  orignh=  99.9 
84   &   rh600964n00\tablenotemark{*}   & 0.78$\pm$0.05 &   5.1   &  20.5  &   2.15 \\ % PGC 53231  orignh=  20.5 
85   &   rh600501n00\tablenotemark{*}   & 3.15$\pm$0.14 &   25.4   &  21.3  &   1.06 \\ % PGC 65001  orignh=  21.3 
     &   rh600718n00   & 1.31$\pm$0.09 &   5.0   &        &   0.44 \\ % PGC 00000  orignh=  99.9 
86   &   rh701300n00\tablenotemark{*}   & 1.36$\pm$0.11 &   6.4   &  20.2  &   2.55 \\ % PGC 69253  orignh=  20.2 
87   &   rh702055n00\tablenotemark{*}   & 2.29$\pm$0.23 &   3.4   &  20.3  &   5.76 \\ % PGC 71031  orignh=  20.3 
\enddata

\tablenotetext{*}{Dataset name used for contour plot in Figures 1$-$15}

\tablenotetext{\ }{
\noindent
Notes on table columns:
(1) IXO catalog number;
(2) archive dataset name;
(3) HRI count rate and error in count rate (from Gaussian plus sloping background model fits);
(4) significance of detection: number of source counts divided by the square root of the
number of background counts in the source detection cell;
(5) Galactic absorption column from the {\sc FTOOLS NH} program (Dickey \& Lockman 1990);
(6) 2$-$10 keV X-ray luminosity estimated from the ROSAT HRI count rate, 
assuming a power-law spectrum with $\Gamma=$1.7 and the Galactic hydrogen column (column 5).
}

\clearpage

\end{deluxetable}

\end{document}